\documentclass{ametsocV6.1}
\usepackage[framed,numbered,autolinebreaks,useliterate]{mcode}
\usepackage{color}
\usepackage{amsmath}

\newcommand{\new}[1]{{\color{black} #1}}

\nolinenumbers

\DeclareMathOperator{\sgn}{sgn}

\begin{document}

\title{Topological Signature of Stratospheric Poincar\'e -- Gravity Waves}

%
%
%

%

\authors{Weixuan Xu\aff{a}\correspondingauthor{Weixuan Xu, weixuan\_xu@brown.edu},
Baylor Fox-Kemper\aff{a},
Jung-Eun Lee\aff{a},
J. B. Marston\aff{b},
Ziyan Zhu\aff{c}}

\affiliation{\aff{a}{Department of Earth, Environment and Planetary Science, Brown University, Providence RI 02912 USA}, \aff{b}{Department of Physics and Brown Theoretical Physics Center, Brown University, Providence, RI 02912-S USA}, \aff{c}{Stanford Institute for Materials and Energy
Sciences, SLAC National Accelerator Laboratory, Menlo Park, CA 94025, USA}}

%
%
 
\abstract{The rotation of the earth breaks time-reversal and reflection symmetries in an opposite sense north and south of the equator, leading to a topological origin for certain atmospheric and oceanic equatorial waves. Away from the equator the rotating shallow water and stably stratified primitive equations exhibit Poincar\'e inertio-gravity waves that have nontrivial topology as evidenced by their strict superinertial timescale and a phase singularity in frequency-wavevector space. This non-trivial topology then predicts, via the principle of bulk-interface correspondence, the existence of two equatorial waves along the equatorial interface, the Kelvin and Yanai waves. To directly test the nontrivial topology of Poincar\'e-gravity waves in observations, we examine ERA5 reanalysis data and study cross-correlations between the wind velocity and geopotential height of the mid-latitude stratosphere at the 50 hPa height.  We find the predicted vortex and anti-vortex in the relative phase of the geopotential height and velocity at the high frequencies of the waves.  By contrast, lower-frequency planetary waves are found to have trivial topology also as expected from theory. These results demonstrate a new way to understand stratospheric waves, and provide a new qualitative tool for the investigation of waves in other components of the climate system.}

\maketitle

%
%
%
%
%
%

%

\section{Introduction}
\label{intro}

Much of what we understand about the climate system is made possible by recognizing the importance of waves in Earth's atmosphere and oceans. Waves are characterized by predictable periodic motion that contrasts with the \new{more irregular or even chaotic} behavior displayed by many other components of the climate system.  Waves are \new{very often} categorized dynamically by matching observed variability to predicted dispersion relations \citep[e.g.,][]{wheeler_convectively_1999,farrar_observations_2008}, 
which relate the frequencies and spatial scales where waves may occur and how they propagate through space and time. The phase of waves receive rather less frequent attention.  Phasing in the meteorological context refers to different disturbances such as fronts and low-level jets coming together in space and time.  Another example in the context of baroclinic instability theory is counterpropagating Rossby waves \citep{Hoskins.1985}. \new{The phase spectrum of convectively generated gravity waves in the tropics has been studied recently using ERA5 reanalysis data \citep{Pahlavan.2023}.}  Some rogue ocean waves may result from the constructive superposition of multiple smaller waves.  
Here we demonstrate a different distinguishing and qualitative feature of certain waves, nontrivial topology, that can also be discerned from observations of relative wave phase.  The nontrivial topology found here in reanalysis data of stratospheric waves agrees with a recent theoretical prediction of \citet{delplace_topological_2017}. 

Remarkably, oceanic and atmospheric waves share fundamental physics with those in quantum matter, and topology plays an important role in the movement of the atmosphere and oceans.  Although the basic equations for idealized equatorial plane waves in the atmosphere and ocean have been long known \citep[see the historical perspective in][]{hendershott_tides_1970} and wavelike variability has been observed in the atmosphere \citep{wheeler_convectively_1999} and ocean \citep{farrar_observations_2008} with the waves linked to phenomena such as El Ni\~no \citep{wyrtki_ninodynamic_1975} and the Madden-Julian Oscillation \citep{madden_detection_1971}, the connection of waves to topology has only recently been discovered \citep{delplace_topological_2017}.  This discovery presents an opportunity to use more sophisticated statistical and mathematical analysis of the wave-like aspects of these phenomena, sharpening our insights into their emergence from diverse variability, fundamental mechanisms, and especially the ability of our modeling systems to appropriately simulate them. 

\begin{figure*}[tbh]
\centering
\noindent\includegraphics[width=.9\textwidth]{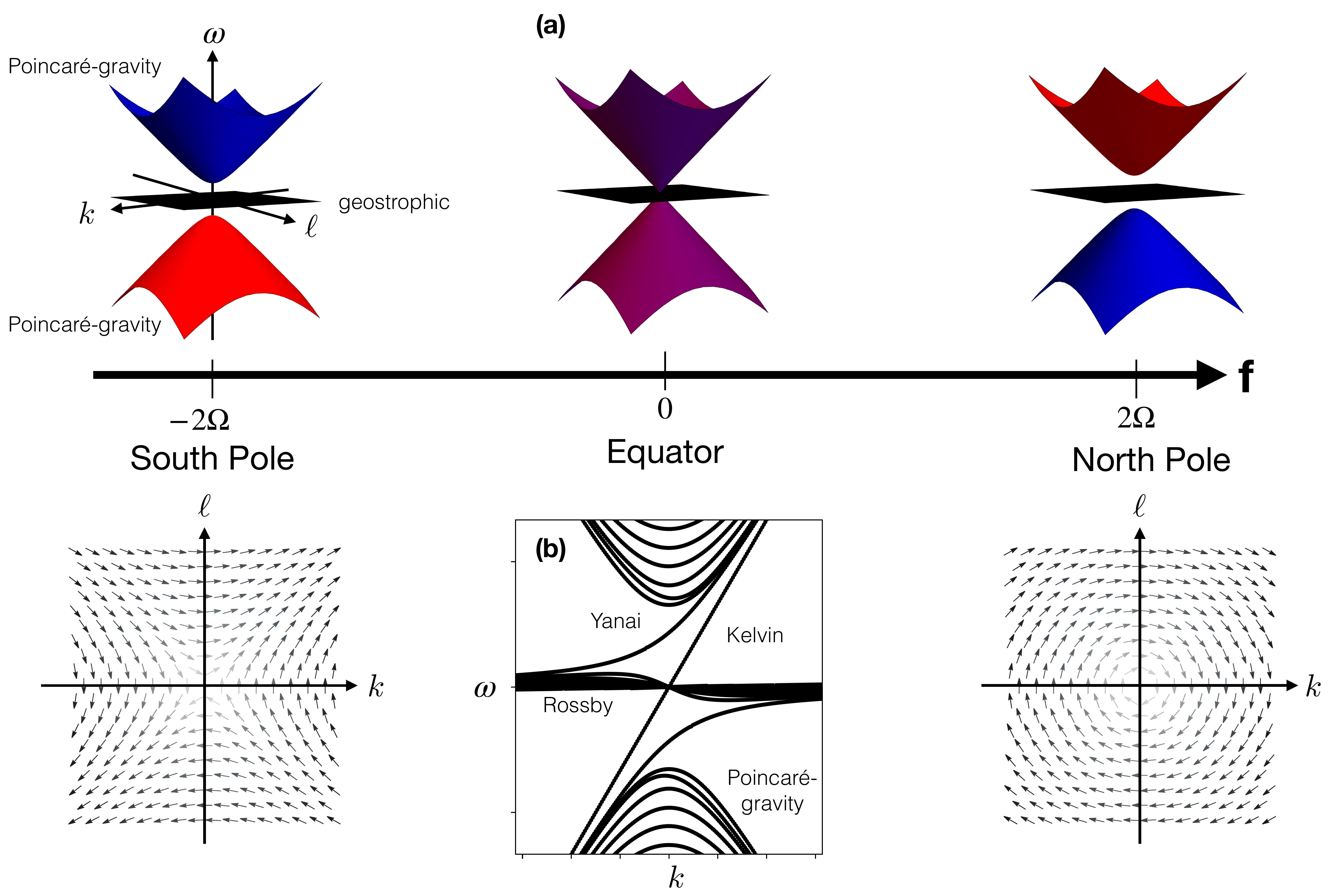}
\caption{
(a) Dispersion relation in frequency-wavevector space for the rotating shallow water equations in the f-plane approximation as a function of latitude. The upper and lower bands are positive and negative superinertial frequency modes of the Poincar\'e waves, and the color indicates the sign of the winding number of the upper band (blue = -1, red = +1) as shown by the plots of the relative phase difference between the height and meridional velocity in the lower half of the figure (argument of $\Xi(k, \ell)$; see text). At the equator $f = 0$, the frequency gap vanishes at a Weyl point, and a topological transition occurs (purple) as the two bands invert.  The subinertial range has only a zero frequency band (black) containing modes in geostrophic balance as there is no $\beta$ effect in the f-plane approximation. The inset (b) shows the equatorial $\beta$-plane dispersion relations with the subinertial quasi-geostrophic Rossby waves, the superinertial Poincar\'e inertio-gravity waves, and the Kelvin and Yanai waves. Vector plots correspond to the southern hemisphere (left) and the northern hemisphere (right) respectively. }
\label{f_concept}
\end{figure*}

Topology is the field of mathematics concerned with the properties of spaces that do not change under continuous deformations. A doughnut and an orange are topologically distinct because the doughnut has a single hole (we say that it has a genus of 1) while the orange has a genus of 0.  Likewise a coffee mug has the doughnut topology because it too has a single hole (the handle).  Another example, more closely connected to the phenomena discussed below, is known as the Hairy Ball or Hedgehog Theorem which says that it is impossible to comb the spines of a hedgehog (because there will always be at least one tuft) \citep{renteln_manifolds_2013}.  By contrast, if hedgehogs had a toroidal shape instead of spherical, their spines could be combed smoothly without any bald spots -- but probably only boring hedgehogs would choose that hairstyle.  

Topology is a powerful tool because it turns \new{certain aspects of} complicated problems into simple ones. \new{Changes in a model or physical system that do not alter the topology leave certain qualitative properties unchanged.} For example, across an interface between topologically distinct states of matter, a general principle known as bulk-boundary or bulk-interface correspondence \citep{hasan_colloquium_2010,Ozawa.2019,Simon.2021} guarantees the existence of boundary or interfacial waves (Fig.~\ref{f_concept} (a)) \new{and these waves persist even when the model or physical system is deformed}. The waves move in one direction and evidence \emph{topological protection}, or immunity to backscattering even in the presence of defects. On a rotating planet, the equator acts as the boundary between two topologically-distinct hemispheres, and so some equatorial waves propagate only in one direction and are topologically protected (Fig.~\ref{f_concept} (b)).  Past uses of topology in fluid mechanics have usually focused on coherent structures in space such as vortices.  Here by contrast we study topology in wavevector-frequency space where it can for instance guarantee the existence of eastward propagating equatorial waves in Earth's climate system \citep{delplace_topological_2017}. In particular there is a topological origin for two well-known equatorially trapped waves, the Kelvin and Yanai modes, caused by the breaking of time-reversal symmetry by Earth's rotation, that helps to explain the robustness of these waves against buffeting by the weather. This resilience may also be implicated in other emergent equatorial wavelike phenomena such as the Madden-Julian Oscillation (MJO), remarkable for its eastward propagation along the equator.  

\new{In this paper we study the topology of waves away from the equatorial region to ascertain whether or not topological signatures can be deduced from observation, and whether or not the signatures found are consistent with theoretical prediction and with bulk-interface correspondence. We first briefly review the theory and refer the reader to background material \citep{hasan_colloquium_2010,delplace_topological_2017,Ozawa.2019,Simon.2021,zhu_topology_2021} for more details.}

Fig. \ref{f_concept}a shows the dispersion relation for the idealized rotating shallow-water model on the f-plane. There are three distinct bands:  Positive and negative frequency Poincar\'e-gravity waves (often referred to as ``inertio-gravity waves'') and a zero-frequency geostrophically balanced mode. The topology of each band is distinct and may be quantified in terms of a winding number (defined below) in frequency-wavevector space \citep{Horsley2022}.  In particular, the Poincar\'e-gravity modes are characterized by a vortex with a winding number of $\pm 1$.  A winding number of $+1$ means that the phase of the complex-valued quantity $\Xi(k, \ell)$ (defined below) increases by $2 \pi$ as one moves around the center of the vortex in a counter-clockwise direction in wavevector space.   The sign of the winding number depends on the sign of the product of the Coriolis parameter and the wave frequency.  Geostrophically balanced modes (that become Rossby waves once the Coriolis parameter is allowed to vary with latitude) are by contrast topologically trivial with zero winding number.  

Evidently the winding number of the Poincar\'e-gravity waves changes by $2$ upon crossing the equator where the Coriolis parameter changes sign; this is known as ``band inversion.''  By bulk-interface correspondence, there must therefore be 2 waves that traverse the otherwise forbidden region of frequency space.  Spectral flow \citep{Faure.2023} in frequency-wavevector space as the zonal wavenumber increases show that the negative frequency Poincar\'e band loses two modes, the geostrophic band gains and loses one mode, and the positive frequency Poincar\'e band gains the two modes. These two modes are the equatorial Kelvin and Yanai (also known as mixed Rossby-gravity) waves.  The two equatorial modes move with an eastward group velocity at all zonal wavenumbers, and this unidirectional propagation reflects the breaking of time-reversal invariance by the planetary rotation.  

The organization of the rest of the paper is as follows.  In Sec. \ref{theory} we review the mathematics of topology in the context of the rotating shallow water equations on the f-plane. Sec. \ref{methods} discusses the atmospheric ERA5 reanalysis data and its processing into spectral space. Topological properties of waves discerned from the data are presented in Sec. \ref{results} and some conclusions are presented in Sec. \ref{conclusions}. The Matlab code that we use is included \new{as supplementary material}.

\section{Theoretical Motivation}
\label{theory}

To motivate our investigation of the topology of waves in the stratosphere we briefly review the linearized rotating shallow water equations on the f-plane following \citet{delplace_topological_2017}. 
\begin{equation}
\begin{cases}
\frac{\partial u}{\partial t}-f_0v=-g\frac{\partial h}{\partial x} \\
\frac{\partial v}{\partial t}+f_0u=-g\frac{\partial h}{\partial y} \\
\frac{\partial h}{\partial t}+H(\frac{\partial u}{\partial x}+\frac{\partial v}{\partial y}) = 0\ .
\end{cases}
\end{equation}
Here $u$ and $v$ are the zonal and meridional velocity, $f_0$ is the Coriolis parameter in the f-plane approximation and $g$ is the gravitational acceleration. Also $h$ is the height anomaly and $H$ is the average depth of the shallow water. Despite the highly idealized nature of these equations, the robustness of topology against continuous deformations means that lessons learned from these equations may be expected to hold in more realistic settings \citep{zhu_topology_2021}.  

By adopting periodic boundary conditions to eliminate any boundaries, the normal modes of the linearized rotating shallow water equations may be easily found by Fourier transformation to frequency-wavevector space followed by diagonalization of a $3 \times 3 $ matrix. Introducing the 3-component vector $\Psi(x, y, t) \equiv (u(x, y, t),~ v(x, y, t),~ h(x, y, t))$ and substituting $\Psi(x, y, t) = \Psi(k, \ell, \omega)~ e^{i(\vec{k}\cdot\vec{x}-\omega t)}$ with $\vec{k} = (k, \ell)$ \new{and $\omega = 2 \pi \nu$ which defines the angular frequency $\omega$ in terms of the frequency $\nu$} we obtain
\begin{equation}
\begin{bmatrix} -i\omega & -f_0 & i g k \\ f_0 & -i\omega & i g \ell \\ i H k & i H \ell & -i\omega \end{bmatrix}~ \left[ \begin{array}{c} u \\ v \\ h \end{array} \right] = 0\ .
\end{equation}
This secular equation is solved to obtain three normal mode angular frequencies. In the long-wavelength limit ($f_0^2 \gg g H (k^2 + \ell^2))$, \new{we recover inertial oscillations of $\omega \approx \pm f_0$ and} the amplitudes of the positive and negative frequency Poincar\'e-gravity waves, \new{and the zero mode,} are:
\begin{equation}
\Psi_{\pm} =\left[ \begin{array}{c} \pm (k \pm i \ell \sgn(f_0)) |f_0| \\ -i (k \pm i \ell \sgn(f_0)) f_0 \\  H(k^2+\ell^2) \end{array} \right],\ \new{\Psi_{0} =\left[ \begin{array}{c} - g \ell \\   g k \\  - i f_0 \end{array} \right]}
\end{equation}
up to an overall normalization.  Note that $\Psi^*_\pm(\vec{k}, f_0) = \Psi_\pm(\vec{k}, -f_0)$ whereas the positive and negative frequency modes are related by band inversion $\Psi_-(\vec{k}, f_0) = \Psi_+(-\vec{k}, -f_0)$. The presence of the $k \pm i \ell \sgn(f_0)$ term \new{for the Poincar\'e-gravity amplitudes} signals the appearance of a phase singularity at the origin in wavevector space. 

Normal modes are defined only up to an overall phase and magnitude.  To quantify their topology we follow \citet{zhu_topology_2021} by introducing a gauge-invariant but complex-valued quantity $\Xi$ defined as follows: 
\begin{eqnarray}
\Xi(k, \ell) \equiv h^*(k, \ell)~ v(k, \ell).
\label{xi}
\end{eqnarray}
We say that $\Xi$ is gauge-invariant because the overall phase of the normal modes cancels out; only the relative phase difference between two components of the waves (in this case, meridional velocity and height) remains. A Fourier transform of $\Xi$ into real space and time reveals it to be a cross-correlation between two field points that are separated in space and time.  As waves propagate simultaneously in both space and time it is perhaps not surprising that non-local correlations play a role in their topological classification.  

In the case of the inertial Poincar\'e-gravity waves the gauge-invariant quantity displays a vortex or antivortex (depending on the signs of the frequency and the Coriolis frequency) centered at the origin in wavevector space:  
\begin{equation}
\Xi_{\pm}(k, \ell) =  -i H f_0 (k^2 + \ell^2)(k \pm i \ell \sgn(f_0))\ .
\label{xi-gravity}
\end{equation}
\new{As noted above, the winding number may be found from the change in phase of $\Xi(k, \ell)$. A winding number of $+1$ ($-1)$ means that the phase of a complex-valued quantity decreases (increases) by $2 \pi$ as one moves around the center of the vortex in a clockwise (counter-clockwise) direction in wavevector space.} In the northern hemisphere, $f_0 > 0$, and the positive-frequency waves have winding number $+1$ whereas negative-frequency waves have winding number $-1$; these topological charges are reversed in the southern hemisphere (the bands invert).  The related product $h^*(k, \ell)~ u(k, \ell)$ involving the zonal velocity exhibits the same winding numbers.
The zero-frequency geostrophic mode, by contrast, has (up to multiplication by a real-valued normalization constant) the following form:
\begin{equation}
\Xi_{0}(k, \ell) = i g f_0 k
\label{Xi_0}
\end{equation}
and thus has a domain wall at $k = 0$ and zero winding number.  Its topological charge therefore vanishes. The same result may be obtained from the barotropic quasigeostrophic equations as the wave height is proportional to the stream function while the meridional velocity is proportional to the zonal derivative of the stream function thus leading in wavevector space to the same form as Eq. \ref{Xi_0}.

\begin{figure*}[tbh]
\centering
\noindent\includegraphics[width=0.7\textwidth]{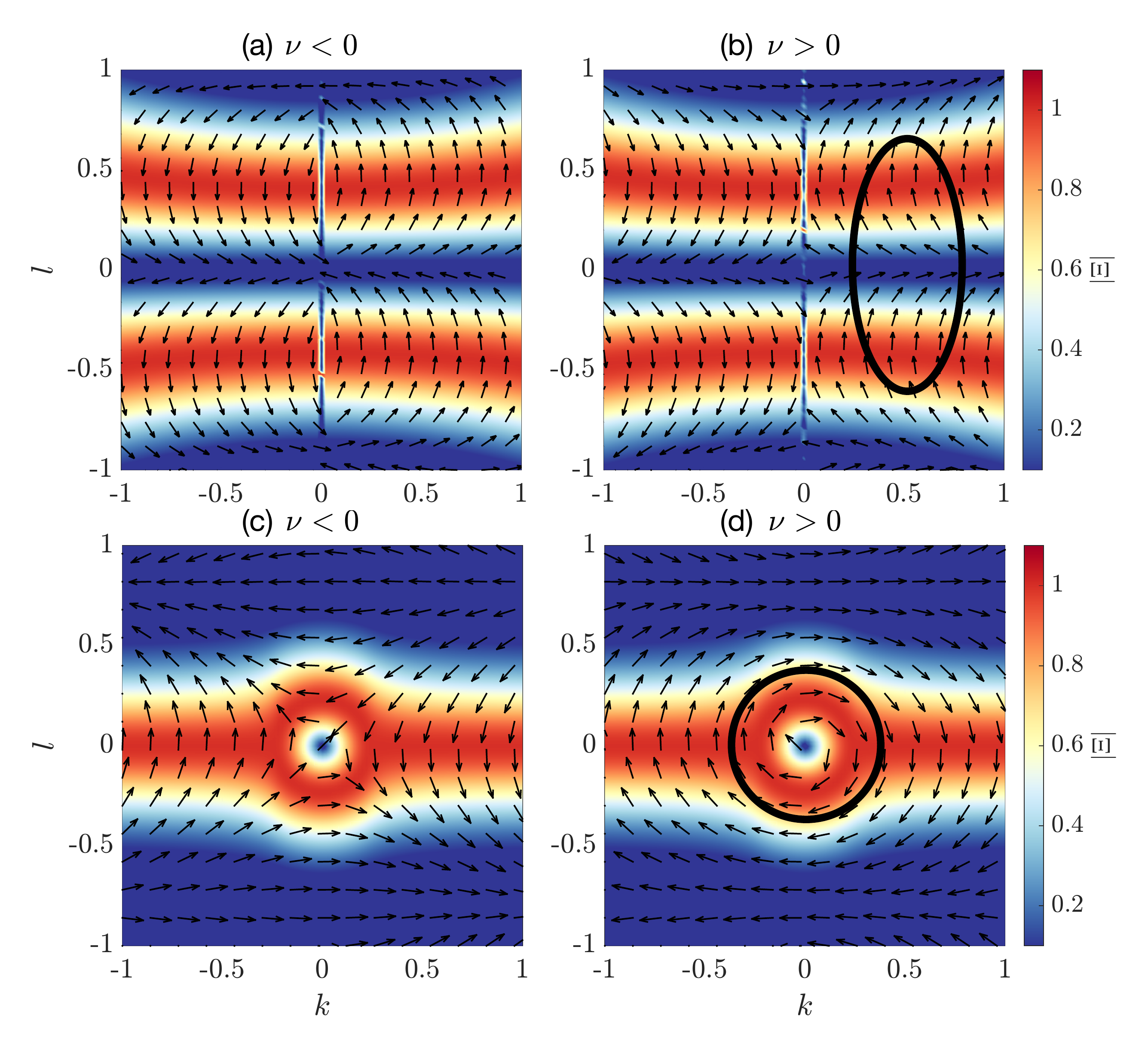}\\
\caption{Theoretical calculation of the cross-correlation $\Xi(k, \ell)$ for normal modes at (a)-(b) low frequencies (Rossby waves) and (c)-(d) high frequencies (Poincar\'e-gravity waves). Here, we perform the calculation for the northern hemisphere with a spatially varying Coriolis parameter to lift the degeneracy of the geostrophic mode: $f(y)=1 + 0.5 \sin( 2\pi y/L_y)$, where $L_y = 10 \pi$ is the domain size. See \citet{zhu_topology_2021} for details. Plots (a) and (c) show the negative frequency modes, and (b) and (d) are for positive frequencies. The color bar represents the magnitude of $\Xi$ and the arrows show the phase of its complex value.  The low-frequency planetary waves have a domain wall at $k = 0$ and are topologically trivial as a closed path that avoids the $\Xi = 0$ domain wall has zero winding number (black oval). While the cross-correlation structure of the geostrophic mode in (a)-(b) depends on the choice of normal modes, the winding number remains zero for all modes. Note that the winding number of the high-frequency Poincar\'e-gravity waves (black circle in the bottom figure) depends on the sign of the frequency of the waves.}
\label{f_theory} 
\end{figure*}

Representing the phase of $\Xi$ with an arrow makes these patterns evident as shown in Figure \ref{f_theory}.  In \citet{delplace_topological_2017} and many other papers, linearized wave equations have been characterized in terms of the topological Chern number.  However the Chern number has several disadvantages.  First, in contrast to systems on spatial lattices (where the Chern number was first applied), for continuous systems \new{the Chern number when calculated as an integral of the Berry curvature} need not be integer and its value depends on how an integral over the Berry curvature is regularized at high wavevectors.  This ambiguity can sometimes be avoided by compactification \citep{delplace_topological_2017,Venaille.2021}.  Here we point out that this is more of a mathematical problem than a physical one because at small scales dissipation is important and ultimately at the smallest scales the fluid description breaks down entirely and is replaced by discrete molecular dynamics.  In any case the ambiguity does not arise for the winding number which is determined at finite wavevectors.  Second, it is unclear how to extend the Chern number to systems with dissipation, driving, or nonlinearities -- all properties of geophysical fluids. Finally it is difficult to compute the Chern number from observations or simulations because it involves an integral of the Berry curvature over wavevector space.  We note that the winding number has recently been utilized in a number of different contexts \citep{Tauber.2019,zhu_topology_2021} including active optical media \citep{Mittal.2016,Ozawa.2019,Simon.2021}.

\section{Methods}
\label{methods}

\new{To investigate the topology of waves from observations by following the approach described above requires a stratified component of the fluid Earth system where observations are made at a sufficiently high sampling rate that Poincar\'e-gravity waves can be diagnosed ($\nu > 1.4$ CPD  in the mid-latitude region) and over a regular spatial grid located away from the equator. Initially we considered the oceans but these requirements are not met by existing observations \citep{crout_physical_2012,lumpkin_noaa_2010}}.  We therefore turn to the stratosphere and avoid for now the troposphere as it is generally less stably stratified and subjected to orographic effects and baroclinic and convective instabilities.  We make use of the ERA5 reanalysis data set which contains high-frequency data sampled evenly over the globe.  Later in the Sec. \ref{conclusions} we sketch a possible way to avoid the requirement for a fine spatial grid of data points.   

\subsection{Reanalysis Data}
\label{data}

\begin{figure*}[tbh]
\noindent\includegraphics[width=\textwidth]{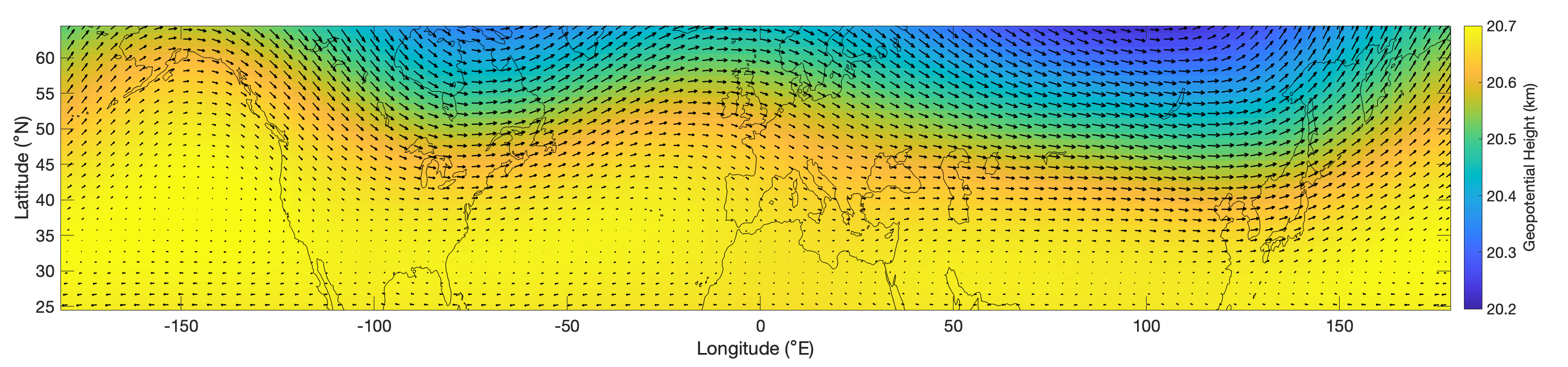}\\
\caption{\new{The geographic area studied in this paper.  Arrows indicate the mean wind at 50 hPa and the color is the mean geopotential height averaged over the 40 year period from 1981 to 2020.}}\label{f_data_spatial_dist} 
\end{figure*}

In order to \new{measure winding numbers} the velocity and geopotential height fields ($u$, $v$, $h$) should satisfy two conditions. First, the region to be studied should be several deformation lengths away from the equator to ensure that the waves are not strongly influenced by equatorially-trapped waves.  In the language of topological physics this is the ``bulk'' region.  Second the data sampling interval should be short enough to resolve signals with a frequency of up to 2  CPD. Following \citet{pahlavan_revisiting_2021} who diagnosed the equatorial Kelvin and Yanai waves as well as Poincar\'e-gravity waves in spectral analysis of the stratosphere, we use ERA5 reanalysis data \citep{hersbach_era5_2020} at the 50 hPa level. \new{Reanalysis data underestimates wave strength \citep{Kim.2015} but topology is robust against changes in strength.}  We sample the horizontal components of the velocity and the geopotential heights at 6-hour intervals over the period of 1981 to 2020.  The spatial domain ranges from 25$^{\circ}$N to 65$^{\circ}$N across all longitudes with a horizontal resolution of 0.25$^{\circ}$, \new{as shown in Fig. \ref{f_data_spatial_dist}. We note that at a latitude of less than 30$^\circ$ there are Poincar\'e-gravity waves of frequency less than 1 CPD but this does not affect our bulk analysis.}

\subsection{Data Processing}
\label{processing}

To work with a manageable data set size, we first reduce the spatial resolution by interpolating the variables to a coarser resolution of 1$^{\circ}$ in latitude and 2.5$^{\circ}$ in longitude.  As the Poincar\'e-gravity waves that we wish to study have long wavelengths this reduction in resolution does not affect the analysis.  Next we remove the seasonal signals from the data by first applying a low-pass boxcar filter with a window of 182 days (half of a year) in the time dimension to isolate the low-frequency component, and then subtract that from data, leaving only the high-frequency components. Poincar\'e-gravity waves appear at frequencies that exceed the local pendulum frequency.  For a latitude of $\phi = 45^\circ$, the frequency is $\nu = 2~ {\rm CPD} *  \sin(\phi) = 1.42$ CPD. 








We follow the spectral analysis of \citet{wheeler_convectively_1999}.  \new{To explore the seasonal dependence of the topological pattern, we cut the time series into non-overlapping segments for each of the different seasons over the 40 year period.}  The use of non-overlapping seasonal segments is justified because the waves of interest have periods much shorter than a season.  Linear trends are removed from each segment, and then tapering windows are applied in the (non-periodic) latitudinal and time dimensions of the data. We use the standard cosine-tapered Tukey window with $\alpha$=0.5 to reduce spectral leakage \citep{prabhu_window_2014}.
\new{After the data has been pre-processed as described above, we perform a 3-dimensional complex discrete Fourier transform on both the geopotential height ($h$) and the horizontal components of the wind velocity ($u$ and $v$), so that cross-correlations can be computed in the frequency-wavevector space.}
\new{The cross-correlation is calculated for each of the} \new{159} \new{time segments and then cross-correlations are computed by averaging over each of the 40 segments for each season} \new{(39 segments for the DJF season).}

The gauge-invariant complex-valued cross-correlation field in frequency - zonal wavenumber - meridional wavenumber space is \new{the product of the} complex-conjugate of the Fourier-transformed geopotential height multiplied by the horizontal components of the wind velocity.  We denote these quantities as $\langle h(\vec{k}; \nu) | u(\vec{k}; \nu) \rangle$ and $\langle h(\vec{k}; \nu) | v(\vec{k}; \nu) \rangle$ where the brackets indicate the processing steps described above.  The complex field $\langle h(\vec{k}; \nu) | v(\vec{k}; \nu) \rangle$ should be compared to the theoretical calculation of $\Xi$ as defined by Eq. \ref{xi} and theory predicts that $\langle h(\vec{k}; \nu) | u(\vec{k}; \nu) \rangle$ will show similar topology. After selecting a particular frequency, the complex-valued data in the two-dimensional $k$-$\ell$ space is averaged over a moving Gaussian spectral window to further improve statistical significance (albeit at the cost of reduced spectral resolution). The Gaussian window here treats the wavevector space pattern as an image and has the standard deviation $\sigma =1.5$ the spacing between points in wavevector space.  Finally the magnitude and phase of the complex-valued fields $\langle h(\vec{k}; \nu) | u(\vec{k}; \nu) \rangle$ and $\langle h(\vec{k}; \nu) | v(\vec{k}; \nu) \rangle$ are visualized with vectors in horizontal wavevector space.  Matlab code that implements this algorithm is presented in the Appendix.


\section{Results}
\label{results}

\begin{figure*}[tbh]
\noindent\includegraphics[width=\textwidth]{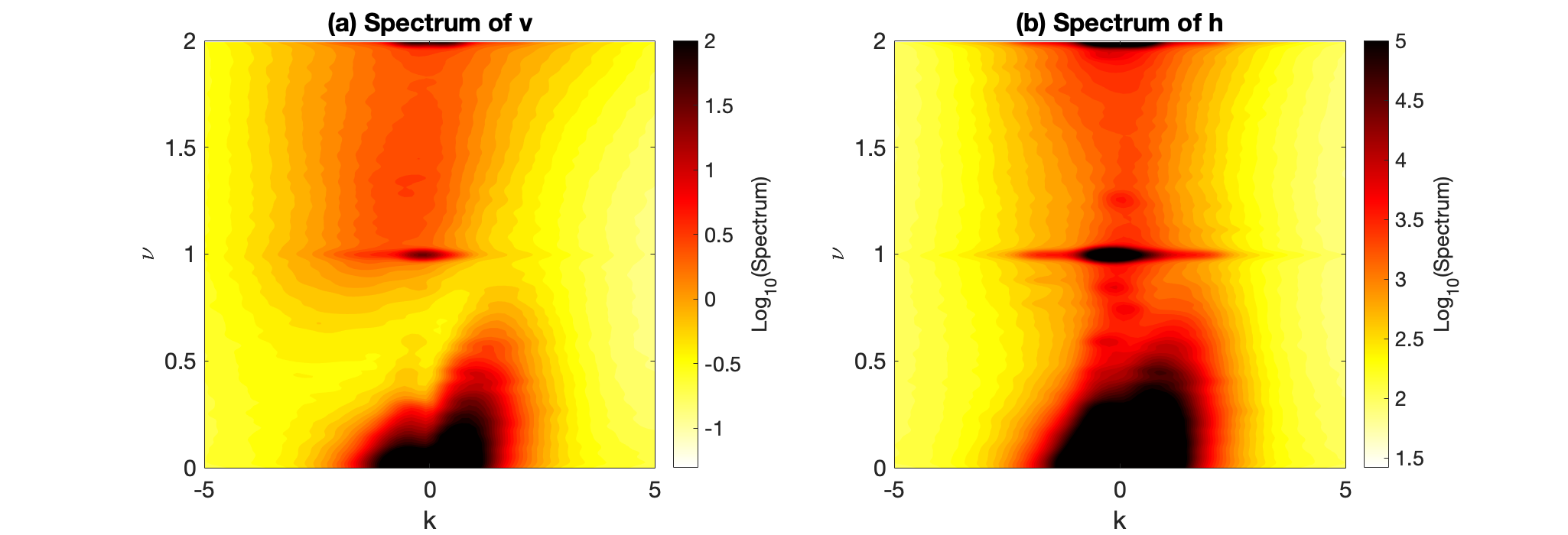}\\
\caption{The 50 hPa power spectra of meridional velocity (v) and geopotential height (h) in the zonal wavenumber ($k$) - frequency ($\nu$) space over latitudes 25$^\circ$N to 65$^\circ$N. The frequency $\nu$ has units of cycles per day (CPD), and $k$ has the unit of cycles per circumference.}\label{f_kxnu}
\end{figure*}

We turn first to the power spectra of the meridional component of the wind velocity and the geopotential height to verify that power can be found in the expected regions of frequency-wavevector space.  We average the power spectra over meridional wavenumbers ($\ell$).  Fig. \ref{f_kxnu} shows the power in zonal wavenumber ($k$) - frequency ($\nu$) space.  Apparent is the sharp signal at $\nu = 1$ CPD due to the diurnal cycle as well as a semi-diurnal harmonic.  High-frequency Poincar\'e-gravity waves are separated by a gap in the power from low-frequency planetary waves as expected, and exhibit more power in the $k < 0$ region \citep{wheeler_convectively_1999}.  The peak in the low-frequency waves at $k > 0$ corresponds to eastward planetary waves whereas part of the signal for $k < 0$ is due to standing waves that contribute equally to both positive and negative zonal wavenumbers. However it is also clear that the spectra are far from the dispersion \new{relation} of the idealized rotating shallow-water \new{waves} shown in Figure \ref{f_concept}.  

\begin{figure*}[tbh]
\noindent\includegraphics[width=\textwidth]{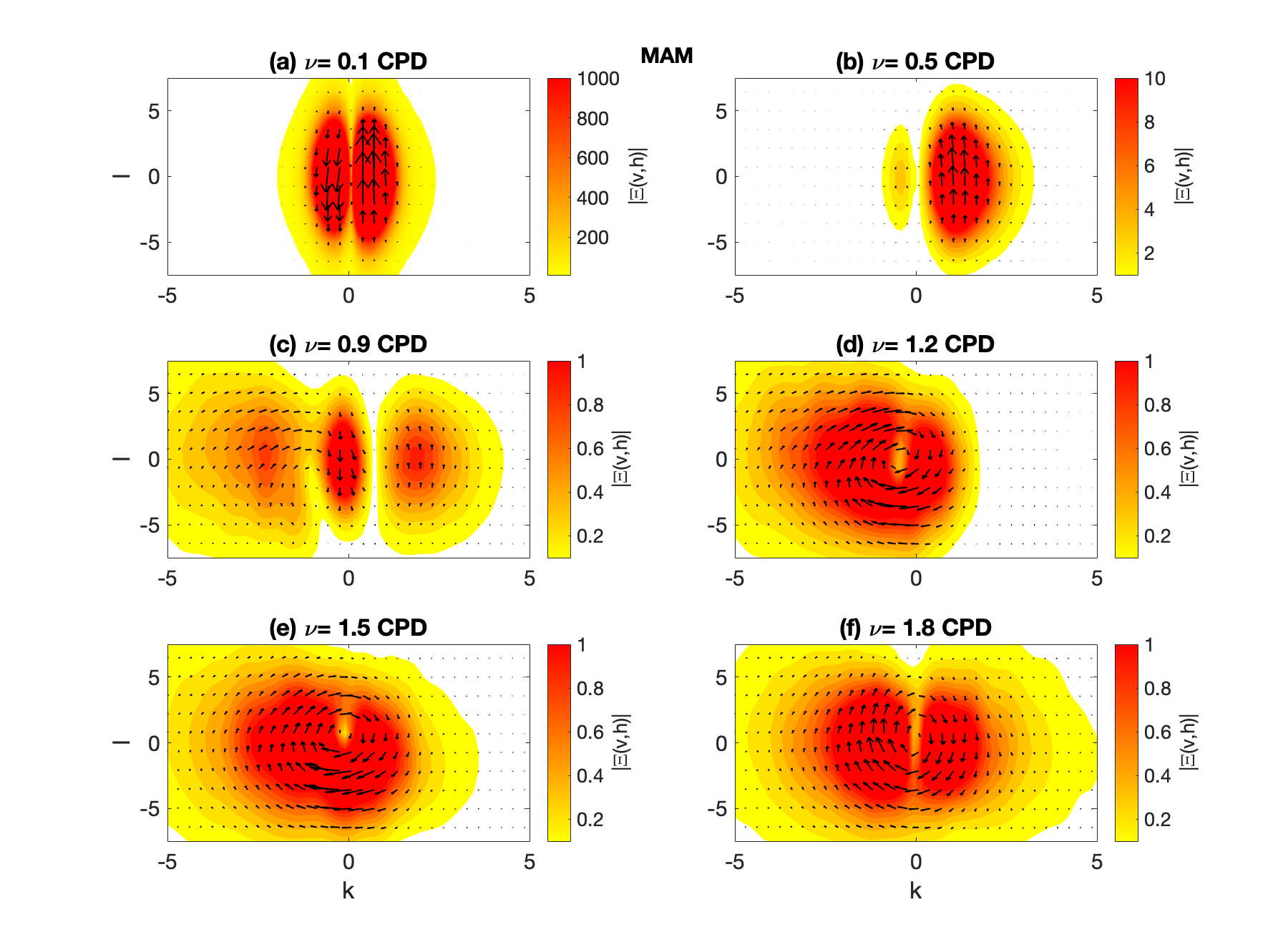}\\
\caption{
The cross-correlation $\langle h(\vec{k}; \nu) | v(\vec{k}; \nu) \rangle$ for the spring season (March, April, May) obtained from ERA5 data for the region 25$^{\circ}$N-65$^{\circ}$N at 50 hPa. Subplots are at different frequencies $\nu$. Colors represent the magnitude $|\langle h(\vec{k}; \nu) | v(\vec{k}; \nu) \rangle|$ on a \new{linear scale} with white = zero, and arrows \new{indicate the complex-value of} $\langle h(\vec{k}; \nu) | v(\vec{k}; \nu) \rangle$.}\label{f_frequencies} 
\end{figure*}

The winding number is discerned by plotting the cross-correlations $\langle h(\vec{k}; \nu) | u(\vec{k}; \nu) \rangle$ and $\langle h(\vec{k}; \nu) | v(\vec{k}; \nu) \rangle$ in $k$-$\ell$ wavevector space. From Fig. \ref{f_theory}, we expect Poincar\'e-gravity waves in the northern hemisphere to have a vortex pattern with winding number = +1 at large positive frequencies.  By contrast, planetary waves will be topologically trivial with a domain wall.  Fig. \ref{f_frequencies} shows that the observational data displays these topological properties.   Vortices with winding numbers equal to $+1$ appear at frequencies larger in magnitude than 1 CPD , consistent with the theoretical expectation for Poincar\'e waves.  
At lower frequencies there are no vortices; instead a topologically-trivial domain wall appears in qualitative agreement with the prediction for Rossby waves, with the phase reversing by $180^\circ$ upon passing through zonal wavenumber $k = 0$.

\begin{figure}[tbh]
\noindent\includegraphics[width=\textwidth]{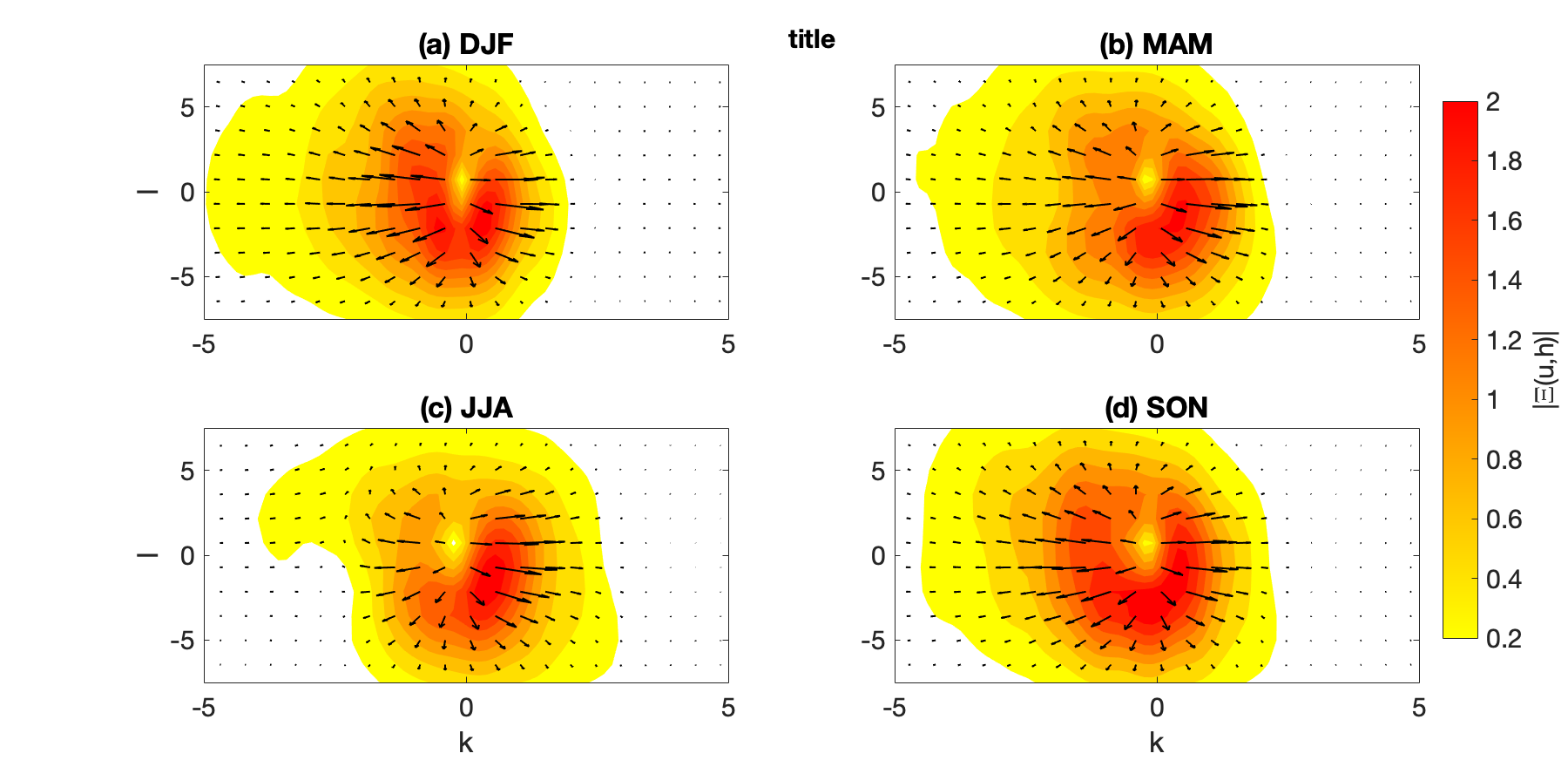}\\
\caption{
Cross-correlation between $\langle h(\vec{k}; \nu) | u(\vec{k}; \nu) \rangle$ obtained from ERA5 data for the region 25$^{\circ}$N - 65$^{\circ}$N at 50 hPa at frequency $\nu$ = 1.5 CPD for the different seasons.}
\label{f_seasons}
\end{figure}

\begin{figure*}[t]
\noindent\includegraphics[width=\textwidth]{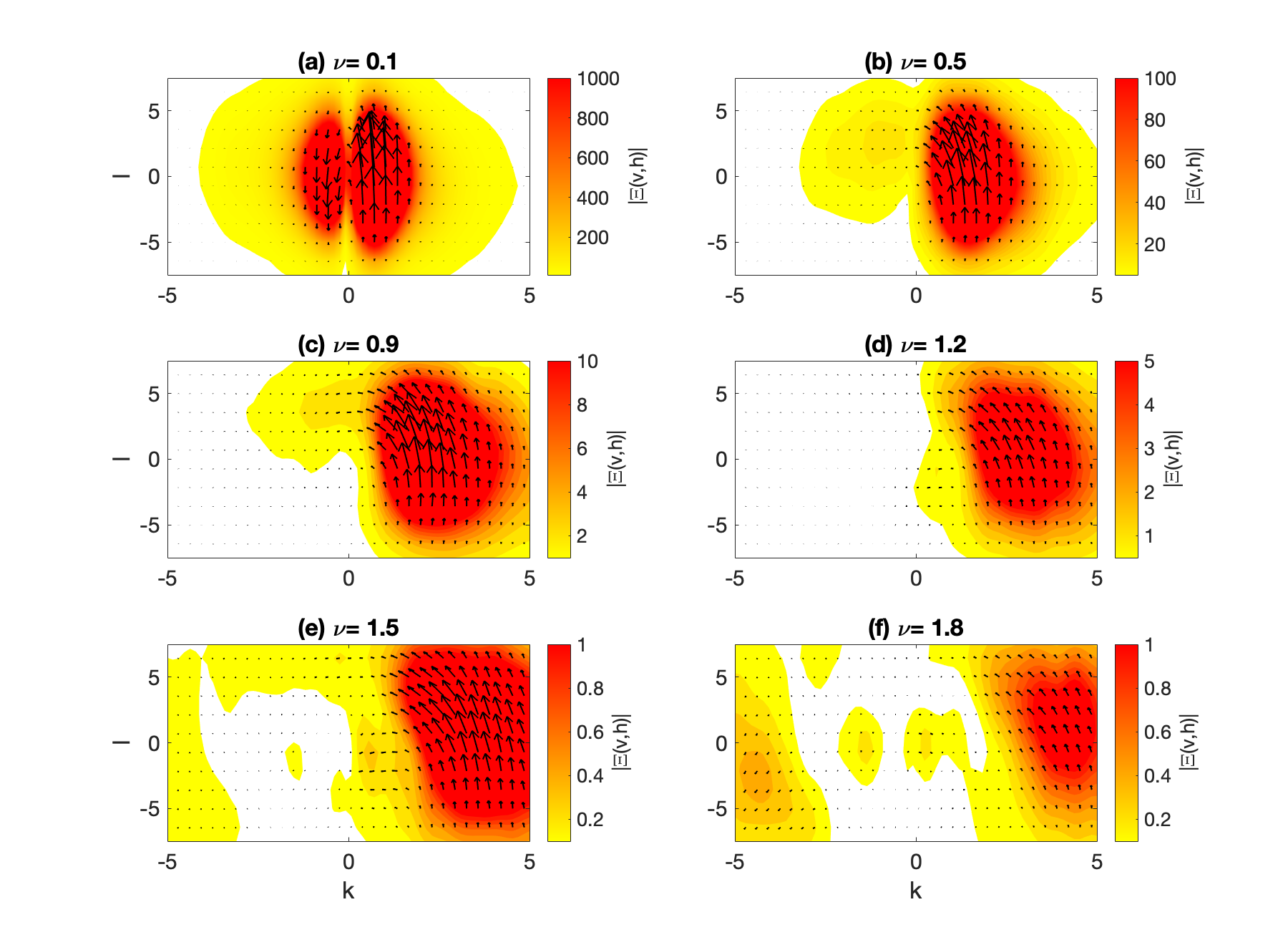}\\
\caption{Same as Fig.\ref{f_frequencies}, but for an altitude of 850 hPa.}\label{f_850hPa}
\end{figure*}

There are some small seasonal variations in the patterns of the Poincar\'e waves (Fig. \ref{f_seasons}). 
We note that the vortex center is offset slightly from the origin in wavevector space.  The offset could be explained by the mean background state flow which varies with season but further investigation is required.
Lower down in the lower troposphere at 850 hPa the Rossby wave pattern persists as expected but the high-frequency patterns are no longer coherent (Fig. \ref{f_850hPa}).  The atmosphere is not strongly stratified and macroturbulence in the form of jets and synoptic scale weather dominates the signal obscuring any gravity waves.  Although vortices are no longer evident, we note that the high-frequency correlations are strongest in magnitude at positive zonal wavenumbers.   

\section{Conclusion}
\label{conclusions}

In this paper we demonstrate that Poincar\'e-gravity waves in the stratosphere have the non-trivial topological signature expected from theoretical prediction.  This work reverses the standard chain of reasoning that begins from the observed dispersion to deduce the existence of equatorial Kelvin and Yanai waves.  From the existence of these waves, the non-trivial topology of the Poincar\'e-gravity waves away from the equator is then inferred from the principle of bulk-interface correspondence and found to agree with theory \citep{delplace_topological_2017}.  Here by contrast we directly interrogate the topology of the superinertial gravity waves away from the equator using ERA5 reanalysis data at 50 hPa pressure level.  We find the theoretically expected winding number of $+1$ at a higher frequency.  The topological signature of Poincar\'e-gravity waves is clear despite the fact that their dispersion relation cannot be clearly discerned in power spectra.  The winding number vanishes at low frequencies and the low-frequency planetary waves have trivial topology that can be plainly distinguished from the Poincar\'e-gravity waves. The non-trivial topological signature of the superinertial gravity waves also disappears in the lower troposphere reflecting the absence of stable stratification.

A shortcoming of the approach described here is that it requires data fields that are well-resolved in both wavevector and (high) frequency space, ruling out the use of spatially sparse data. However it may be possible to work directly in real space instead of wavevector space (while remaining in frequency space):  A spatial Fourier transform of $\Xi_\pm(k, \ell)$ (Eq. \ref{xi-gravity}) shows that $\Xi_\pm(x, y)$ too displays a vortex.  It may be possible to detect the vortex in spatially-sparse buoy \new{or balloon} data provided that the sampling rate is sufficiently high to capture superinertial frequencies.  We leave this for future work. 

The mathematics of topology has great predictive power because it makes complicated problems simple by focusing on robust features.  The topology we investigate here plays out in frequency-wavevector space, rather than in real space.  Topology in this setting is a new tool for climate science that is relatively immune to background noise as the distinct signatures of topology found in ERA5 observations of Poincar\'e-gravity waves qualitatively distinguish them from planetary waves. \new{We show for the first time that Poincar\'e-gravity waves in the stratosphere have a striking non-trivial topology as evidenced by a winding number of $\pm 1$.}  The non-trivial topology of the waves implies the existence of Kelvin and Yanai waves in the stratosphere that have been shown to be a component of the Quasi-Biennial Oscillation \citep{pahlavan_revisiting_2021}.  Future application to other emergent wavelike phenomena such as the Madden-Julian Oscillation may be envisioned.    

\clearpage
\acknowledgments
We thank Tamara Barriquand, George Kiladis, Dung Nguyen, JP O'Brien, and Antoine Venaille for helpful discussions.  This work was supported in part by a grant from the Institute at Brown for Environment and Society and by a grant from the Simons Foundation (Grant number 662962, GF).  It was performed in part at the Aspen Center for Physics, which is supported by National Science Foundation grant PHY-2210452. Z.Z. is supported by a Stanford Science fellowship.

%
%
\datastatement
The ERA5 reanalysis wind and geopotential height data is provided by the Copernicus Climate Change Service (C3S) at ECMWF (https://www.ecmwf.int/en/forecasts/dataset/ecmwf-reanalysis-v5, accessed on 14 April 2022).

%




\vfill\eject


\centerline{Appendix}

\centerline{MATLAB Code for Cross-Correlations in Spectral Space}

\begin{lstlisting}
% %%%%%%%%%%%%   Read Data  %%%%%%%%%%%%
clear
fname='V_ERA.nc';
readncfile
fname='Z_ERA.nc';
readncfile

% %%%%%%%%%%%%   Lower Spatial Resolution  %%%%%%%%%%%%
lat1=linspace(25,65,41).';
long1=linspace(-180,180,145).';

% %%
[X,Y]=meshgrid(longitude,latitude);
[X1,Y1]=meshgrid(long1(1:144),lat1(1:40));

% %%
for i=1:length(time)
     v1(:,:,i)=interp2(X,Y,squeeze(v(:,:,i)).',X1,Y1);
     z1(:,:,i)=interp2(X,Y,squeeze(z(:,:,i)).',X1,Y1);
 end

% %%%%%%%%%%%%   Filter Out Low-Frequency Signals  %%%%%%%%%%%%
 for i=1:40
     i
     for j=1:144
         cv=squeeze(v1(i,j,:));
         fx_v=vfilt(cv,ones(182,1)./182,'mirror');
         resv=cv-fx_v;
         vano(i,j,:)=resv;

         cz=squeeze(z1(i,j,:));
         fx_z=vfilt(cz,ones(182,1)./182,'mirror');
         resz=cz-fx_z;
         zano(i,j,:)=resz;
     end
 end

% %%%%%%%%%%%%  Divide into Segments of seasons  %%%%%%%%%%%%
monthday=[31,28,31,30,31,30,31,31,30,31,30,31];
yearday=cumsum(monthday);

for i=1:40
     segment_v(i,:,:,:)=v1(:,:,(i-1)*365*4+yearday(2)*4+1:(i-1)*365*4+yearday(5)*4);
     segment_z(i,:,:,:)=z1(:,:,(i-1)*365*4+yearday(2)*4+1:(i-1)*365*4+yearday(5)*4);
end % here is MAM for example

[k1,k2,k3,k4]=size(segment_v);
% %%

% %%%%%%%%%%%%  Taper  %%%%%%%%%%%%
% %% Tapering in time dimension %%
time1=1/4:1/4:92;
n=length(time1);
w=tukeywin(n);
for i=1:k1
     for j=1:k2
         for k=1:k3
             y_v=squeeze(segment_v(i,j,k,:));
             b_v=time1.'\y_v;
             ymean_v=time1.'*b_v;
             vanod(i,j,k,:)=(y_v-ymean_v).*w;

             y_z=squeeze(segment_z(i,j,k,:));
             b_z=time1.'\y_z;
             ymean_z=time1.'*b_z;
             zanod(i,j,k,:)=(y_z-ymean_z).*w;
         end
     end
 end
% %%
% %% Tapering in meridional dimension %%
nlat=k2;
w1=tukeywin(nlat);
for i=1:k1
     i
     for j=1:k4
         for k=1:k3
             vanod1(i,:,k,j)=squeeze(vanod(i,:,k,j)).'.*w1;
             zanod1(i,:,k,j)=squeeze(zanod(i,:,k,j)).'.*w1;
         end
     end
 end

% %%
% %%%%%%%%%%%%  3-D Fourier Transform  %%%%%%%%%%%%
 for i=1:k1
     spec_v(i,:,:,:)=fftshift(fftn(squeeze(vanod1(i,:,:,:))))/k2/k3/k4;
     spec_z(i,:,:,:)=fftshift(fftn(squeeze(zanod1(i,:,:,:))))/k2/k3/k4;
 end

% %%
% %%%%%%%%%%%%  Calculate Cross-Spectrum  %%%%%%%%%%%%
 pi=3.1415926536;
 dt=1/4;
 dx=2*pi/144;
 dy=pi/180;

 df=1./k4./dt;
 dkx=1./k3./dx;
 dky=1./k2./dy;

 spec_cross=spec_v.*conj(spec_z)./df./dkx./dky;

% %%%%%%%%%%%%%%  Visualize  %%%%%%%%%%%%
f=[-fliplr(1:(k4/2)) 0 (1:(k4/2-1))].*df;

 kx=[-fliplr(1:(k3/2))+0.5 (1:(k3/2))-0.5].*dkx;
 ky=[-fliplr(1:(k2/2))+0.5 (1:(k2/2))-0.5].*dky;

% %% Find corresponding frequencies, sum up segments %%
 crossxy_1=squeeze(sum(spec_cross(:,:,:,176),1)); 
 crossxy_5=squeeze(sum(spec_cross(:,:,:,139),1)); 
 crossxy_9=squeeze(sum(spec_cross(:,:,:,102),1));
 crossxy_12=squeeze(sum(spec_cross(:,:,:,75),1));
 crossxy_15=squeeze(sum(spec_cross(:,:,:,47),1)); 
 crossxy_18=squeeze(sum(spec_cross(:,:,:,19),1));


 h = fspecial('gaussian',[k2 k3],1.5);
 result_1 = imfilter(crossxy_1, h, 'replicate');
 result_5 = imfilter(crossxy_5, h, 'replicate');
 result_9 = imfilter(crossxy_9, h, 'replicate');
 result_12 = imfilter(crossxy_12, h, 'replicate');
 result_15 = imfilter(crossxy_15, h, 'replicate');
 result_18 = imfilter(crossxy_18, h, 'replicate');


 gapkx=2;
 gapky=1;
 kx1=kx(1:gapkx:k3);
 ky1=ky(1:gapky:k2);
 result1_1=result_1(1:gapky:k2,1:gapkx:k3);
 result1_5=result_5(1:gapky:k2,1:gapkx:k3);
 result1_9=result_9(1:gapky:k2,1:gapkx:k3);
 result1_12=result_12(1:gapky:k2,1:gapkx:k3);
 result1_15=result_15(1:gapky:k2,1:gapkx:k3);
 result1_18=result_18(1:gapky:k2,1:gapkx:k3);

 avgphi3_1=angle(result1_1);
 avgphi3_5=angle(result1_5);
 avgphi3_9=angle(result1_9);
 avgphi3_12=angle(result1_12);
 avgphi3_15=angle(result1_15);
 avgphi3_18=angle(result1_18);


% %%%%%%%%%%%%%%%%%%%%%%%%%%%%%%%%%%%%%%%%%%%%%

 f=figure;
 f.Position(3:4) = [900 660];
 quiversize=0.6;
 color='k';

 xlimrange=[-5 5];
 ylimrange=[-7.5 7.5];

 subplot(3,2,1)
 clims=-1:0.1:3;
 contourf(kx,ky,log10(abs(result_1)),clims,'LineStyle','none');
 %,clims
 title('(a) \nu= 0.1 CPD')
 xlim(xlimrange)
 ylim(ylimrange)
 ylabel('l')
 set(gca,'FontSize',15)
 colormap(flipud(autumn))
 hold on

 quiver(kx1,ky1,cos(avgphi3_1),sin(avgphi3_1),quiversize,color,'LineWidth',1)

% %%%%%%%%%%%%%%%%%%%%%%%%%%%%%%%%%%%%%%%%%%
 subplot(3,2,2)
 contourf(kx,ky,log10(abs(result_5)),clims,'LineStyle','none');
 %,clims
 title('(b) \nu= 0.5 CPD')
 xlim(xlimrange)
 ylim(ylimrange)
 set(gca,'FontSize',15)
 colormap(flipud(autumn))
 hold on

 quiver(kx1,ky1,cos(avgphi3_5),sin(avgphi3_5),quiversize,color,'LineWidth',1)

% %%%%%%%%%%%%%%%%%%%%%%%%%%%%%%%%%%%%%%%%%%

 subplot(3,2,3)
 contourf(kx,ky,log10(abs(result_9)),clims,'LineStyle','none');
 %,clims
 title('(c) \nu= 0.9 CPD')
 xlim(xlimrange)
 ylim(ylimrange)
 ylabel('l')
 set(gca,'FontSize',15)
 colormap(flipud(autumn))
 hold on

 quiver(kx1,ky1,cos(avgphi3_9),sin(avgphi3_9),quiversize,color,'LineWidth',1)

% %%%%%%%%%%%%%%%%%%%%%%%%%%%%%%%%%%%%%%%%%%%

 subplot(3,2,4)
 contourf(kx,ky,log10(abs(result_12)),clims,'LineStyle','none');
 %,clims
 title('(d) \nu= 1.2 CPD')
 xlim(xlimrange)
 ylim(ylimrange)
 set(gca,'FontSize',15)
 colormap(flipud(autumn))
 hold on

 quiver(kx1,ky1,cos(avgphi3_12),sin(avgphi3_12),quiversize,color,'LineWidth',1)

% %%%%%%%%%%%%%%%%%%%%%%%%%%%%%%%%%%%%%%%%%%%%
 subplot(3,2,5)
 contourf(kx,ky,log10(abs(result_15)),clims,'LineStyle','none');
 %,clims
 title('(e) \nu= 1.5 CPD')
 xlim(xlimrange)
 ylim(ylimrange)
 xlabel('k')
 ylabel('l')
 set(gca,'FontSize',15)
 colormap(flipud(autumn))
 hold on

 quiver(kx1,ky1,cos(avgphi3_15),sin(avgphi3_15),quiversize,color,'LineWidth',1)

% %%%%%%%%%%%%%%%%%%%%%%%%%%%%%%%%%%%%%%%%%%%

 subplot(3,2,6)
 contourf(kx,ky,log10(abs(result_18)),clims,'LineStyle','none');
 %,clims
 title('(f) \nu= 1.8 CPD')
 xlim(xlimrange)
 ylim(ylimrange)
 xlabel('k')
 set(gca,'FontSize',15)
 colormap(flipud(autumn))
 hold on

 quiver(kx1,ky1,cos(avgphi3_18),sin(avgphi3_18),quiversize,color,'LineWidth',1)

 \end{lstlisting}

%



\end{document}